\documentclass[twocolumn]{aa}  

\usepackage{graphicx}
\usepackage{txfonts}
\usepackage{lscape}
\usepackage{booktabs}
\usepackage{wasysym}
\usepackage{arydshln}
\usepackage[bookmarks=false,colorlinks=true,linkcolor=black,citecolor=cyan,filecolor=black,urlcolor=cyan]{hyperref}
\usepackage{array}

\newcolumntype{C}[1]{>{\centering\arraybackslash}p{#1}}

\begin{document} 
	
	\title{The complex time and energy evolution of quasi-periodic eruptions in eRO-QPE1}
	
	\author{R. Arcodia
		\inst{1},
		G. Miniutti
		\inst{2},
		G. Ponti
		\inst{3, 1},
		J. Buchner
		\inst{1},
		M. Giustini
		\inst{2},
		A. Merloni
		\inst{1},
		K. Nandra
		\inst{1},
		F. Vincentelli
		\inst{4},
		E. Kara
		\inst{5},
		M. Salvato
		\inst{1},
		\and
		D. Pasham
		\inst{5}
		}
	
	\institute{Max-Planck-Institut f\"ur extraterrestrische Physik (MPE), Giessenbachstrasse 1, 85748 Garching bei M\"unchen, Germany\\
		\email{arcodia@mpe.mpg.de}
		\and
	Centro de Astrobiología (CSIC-INTA), ESAC campus, 28692 Villanueva de la Cañada, Madrid, Spain
	\and
	INAF-Osservatorio Astronomico di Brera, Via Bianchi 46, 23807 Merate, LC, Italy
	\and
	Villanova University, Department of Physics, Villanova, PA 19085, USA
	\and
	MIT Kavli Institute for Astrophysics and Space Research, 70 Vassar Street, Cambridge, MA 02139, USA
	}
	
	\date{Received ; accepted }
	
	
	\abstract{Quasi-periodic eruptions (QPEs) are recurrent X-ray bursts found so far in the nuclei of low-mass galaxies. Their trigger mechanism is still unknown, but recent models involving one or two stellar-mass companions around the central massive ($\approx10^5-10^6M_{\astrosun}$) black hole have gathered significant attention. While these have been compared only qualitatively with observations, the phenomenology of QPEs is developing at a fast pace, with the potential to reveal new insights. Here we report two new observational results found in eRO-QPE1, the brightest QPE source discovered so far: i) the eruptions in eRO-QPE1 occur sometimes as single isolated bursts, and at others as chaotic mixtures of multiple overlapping bursts with very different amplitudes; ii) we confirm that QPEs peak at later times and are broader at lower energies, with respect to higher energies while, for the first time, we find that QPEs also start earlier at lower energies. Furthermore, eruptions appear to undergo an anti-clockwise hysteresis cycle in a plane of hardness ratio versus total count rate. Behavior i) was not found before in any other QPE source and implies that if a common trigger mechanism is in place for all QPEs, it must be able to produce both types of timing properties, regular and complex. Result ii) implies that the X-ray emitting component does not have an achromatic evolution even during the start of QPEs, and that the rise is harder than the decay at a given total count rate. This specific energy dependence could be qualitatively compatible with inward radial propagation during the rise within a compact accretion flow, the presence of which is suggested by the stable quiescence spectrum observed in general for QPE sources.}
	\keywords{}
	
	\titlerunning{Evolution of quasi-periodic eruptions in eRO-QPE1}
	\authorrunning{R. Arcodia et al.} 
	\maketitle

	%
	\defcitealias{Miniutti+2019:qpe1}{M19}
	\defcitealias{Giustini+2020:qpe2}{G20}
	\defcitealias{Arcodia+2021:eroqpes}{A21}
	\section{Introduction}	

\begin{figure}[tb]
		\centering
		\includegraphics[width=0.89\columnwidth]{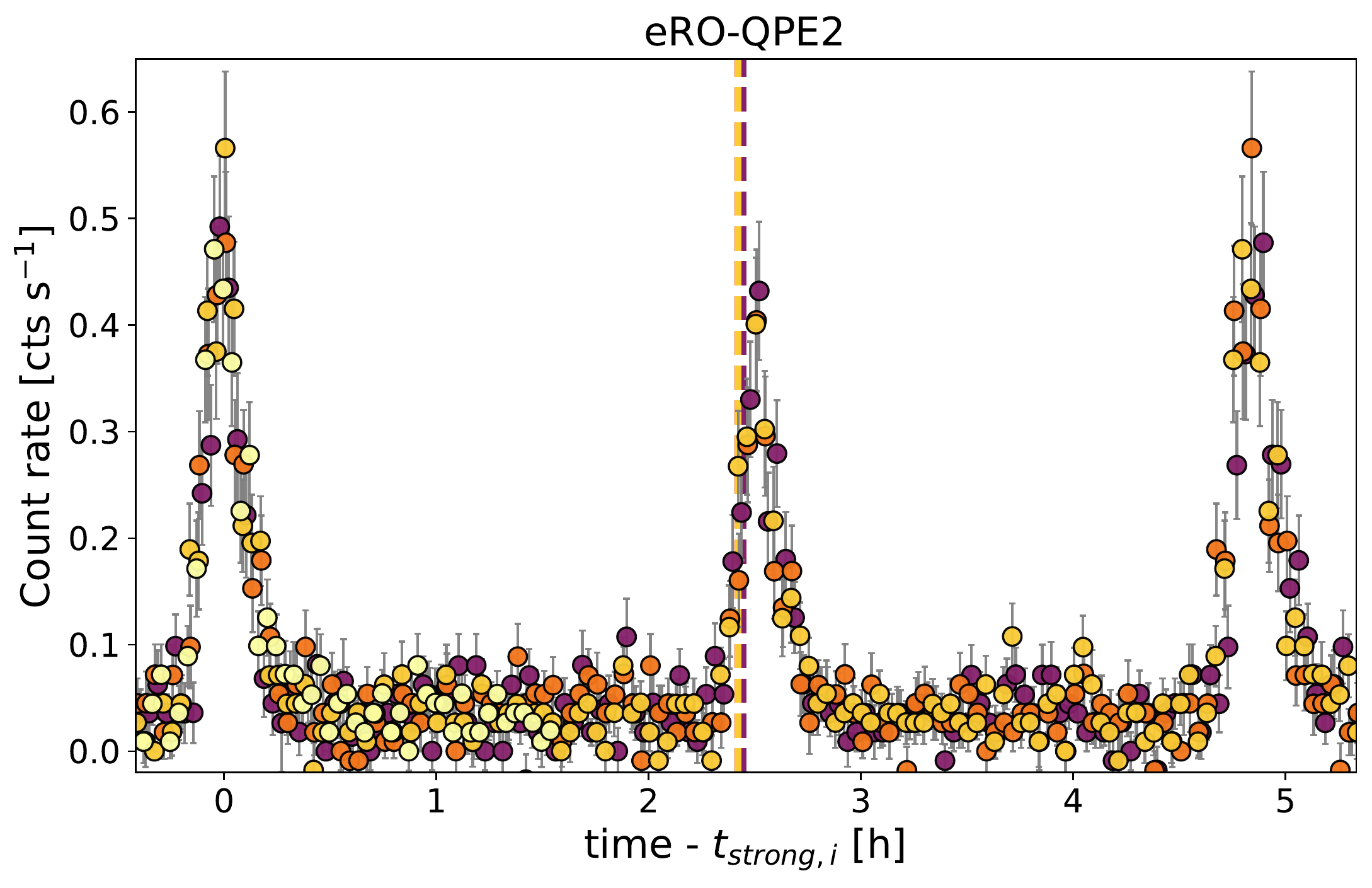}
		\includegraphics[width=0.89\columnwidth]{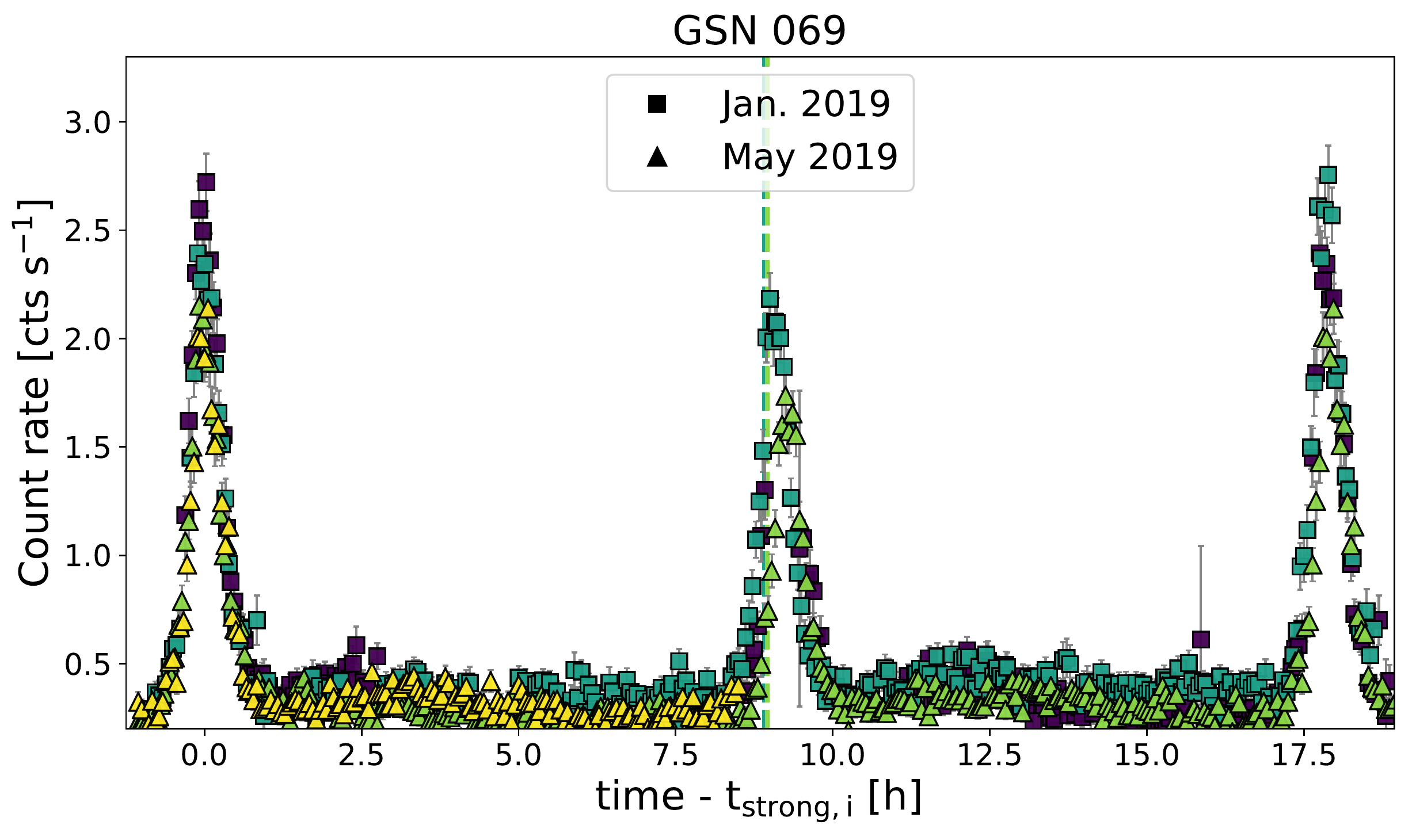}		\includegraphics[width=0.89\columnwidth]{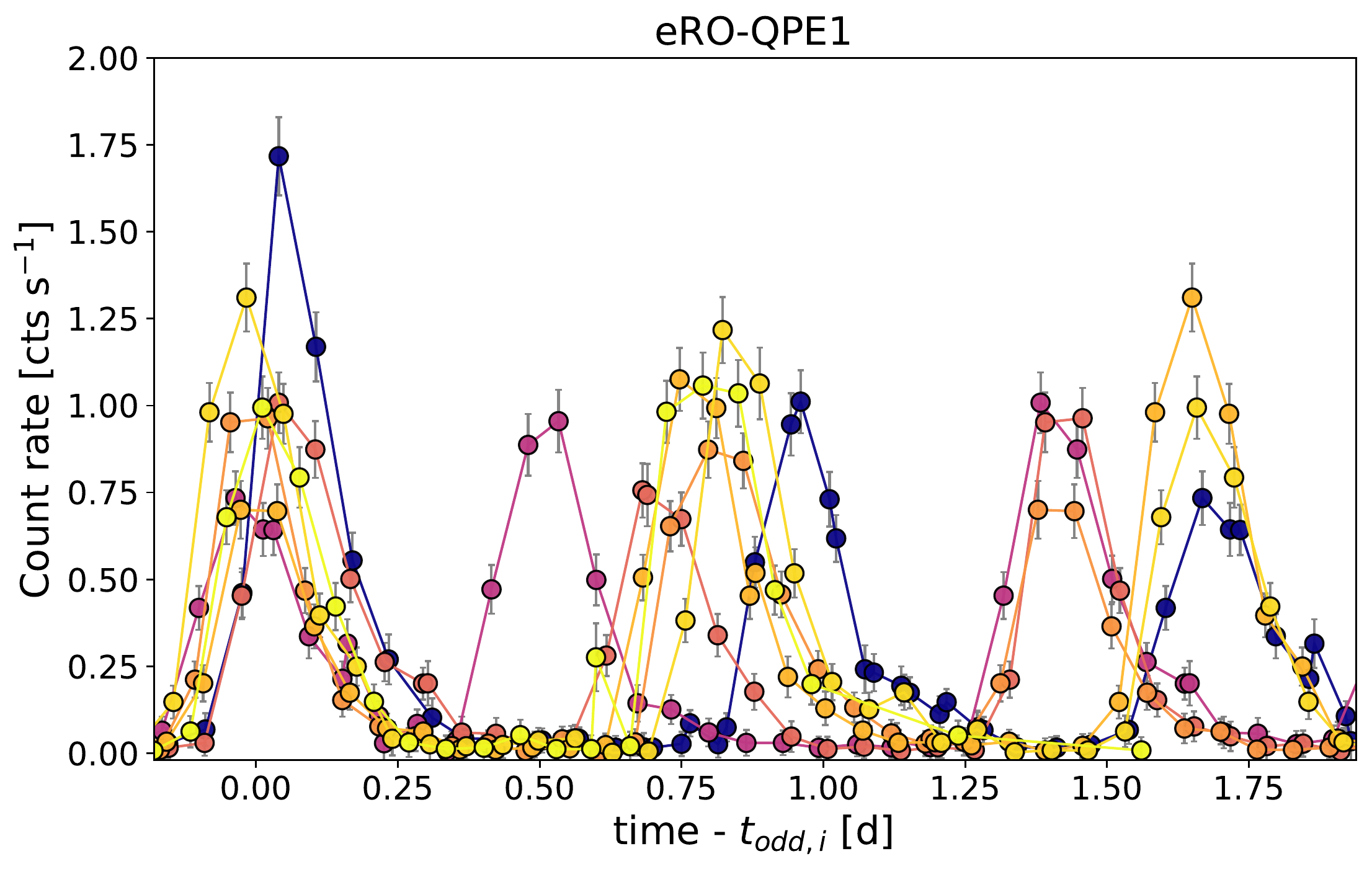}
		\caption{X-ray eruptions of the sources eRO-QPE2 (top, refer to \citetalias{Arcodia+2021:eroqpes}) and GSN\,069 (middle, refer to \citetalias{Miniutti+2019:qpe1}), compared to those of eRO-QPE1 (bottom, refer to \citetalias{Arcodia+2021:eroqpes}). The first two sources show an alternation of longer and shorter recurrence times and stronger and weaker bursts. In these plots (top and middle), eruptions of each source are overlapped in time-space by scaling each strong-weak-strong cycle (shown with a different color) with the related peak time of the first strong burst. Dashed vertical lines show the time corresponding to half of the separation between two strong bursts (i.e. between consecutive bursts aligned to zero), highlighting that the intermediate weak bursts occur systematically later in time (and that two recurrence times alternate). 
		In the bottom panel, the full light curve of eRO-QPE1 reported in \citetalias{Arcodia+2021:eroqpes} is here separated in chunks of two consecutive cycles and scaled in time at the peak of the odd eruptions, to highlight a more complex distribution of recurrence and burst duration times.}
		\label{fig:qpe2_lc}
	\end{figure}

	A very peculiar and novel entry in the catalog of X-ray emitting sources are the so-called X-ray quasi-periodic eruptions (QPEs), so far comprising a handful of sources: GSN\,069 \citep[][hereafter \citetalias{Miniutti+2019:qpe1}]{Miniutti+2019:qpe1}, RX J1301.9+2747 \citep[][hereafter \citetalias{Giustini+2020:qpe2}]{Giustini+2020:qpe2}, eRO-QPE1 and eRO-QPE2  \citep[][hereafter \citetalias{Arcodia+2021:eroqpes}]{Arcodia+2021:eroqpes} and, possibly\footnote{We refer to this as a candidate because only 1.5 QPE-like flares were detected. However, there are many similarities between its observational properties and those of the other known QPE sources.}, XMMSL1 J024916.6-041244 \citep{Chakraborty+2021:qpe5cand}. These soft X-ray eruptions repeat quasi-periodically every few hours and show a count rate increase of more than one order of magnitude over a quiescence level. Despite some scatter in the arrival times, QPEs have so far appeared as orderly pulses (see an example in Fig.~\ref{fig:qpe2_lc}) separated by a steady plateau, usually detected at $\approx0.3-1.6\times10^{41}$\,erg\,s$^{-1}$ in the soft X-ray band (\citetalias{Miniutti+2019:qpe1,Giustini+2020:qpe2,Arcodia+2021:eroqpes}; \citealp{Chakraborty+2021:qpe5cand}).
	
	QPE have so far been identified in nearby low-mass galaxies \citepalias[with
	stellar masses of $M_{star}\approx1-\text{few}\times10^9\,M_{\astrosun}$;][]{Arcodia+2021:eroqpes}. These galaxies host black holes of $M_{BH}\approx10^5-\text{few}\times10^6\,M_{\astrosun}$, as inferred - with the usual large uncertainties - from methods which are both independent (\citealp{Shu+2017:Rx}; \citetalias{Miniutti+2019:qpe1,Giustini+2020:qpe2}; \citealp{Wevers+2022:hosts}) and dependent on the stellar mass (\citealp{Strotjohann+2016:massqpe5}; \citetalias{Arcodia+2021:eroqpes}). Given their peak X-ray luminosity ($L_{0.5-2.0\,keV}\approx10^{42}-10^{43}$\,erg\,s$^{-1}$), the above black hole mass estimates suggest that QPEs peaks are close to being Eddington limited. Studies of the QPEs' host galaxies have found them to be both active (\citealp{Esquej+2007:galqpe5}; \citetalias{Miniutti+2019:qpe1,Giustini+2020:qpe2}) and, quite surprisingly, seemingly inactive galaxies \citepalias{Arcodia+2021:eroqpes}. However, recent studies performing a careful subtraction of the host galaxy's stellar population have found evidence of an ionizing source in addition to star formation for the narrow lines in all five QPEs \citep{Wevers+2022:hosts}. Nevertheless, the absence of broad lines in the optical spectra and of infrared photometry excess associated with the dusty environment typical of active galactic nuclei (AGN), intriguingly suggests that a pre-existing canonical AGN-like accretion flow is not a pre-requisite for QPEs. These fascinating objects might therefore provide a new channel to study how massive black holes are activated in the nuclei of low-mass galaxies through transient accretion events, which is, so far, a poorly studied regime in the black-hole galaxy co-evolution history \citep[e.g.,][]{Kormendy+13:coev,Heckman+2014:bhcoev,Reines+2015:bhm_gal}.
	
	The origin of QPEs is still unclear. Proposed scenarios so far include some types of instability in the accretion disk (\citetalias{Miniutti+2019:qpe1}; \citealp{Sniegowska+2020:instab,Raj+2021:disk_tearing}) and gravitational lensing in a black hole binary of mass-ratio close to unity \citep{Ingram+2021:qpes_lensing}, though both are currently disfavored (\citealp{Ingram+2021:qpes_lensing}; \citetalias{Arcodia+2021:eroqpes}). Lately, models based on a two or three body system with a massive black hole and at least one stellar-mass companion have gained significant attention (\citealp{King2020:gsn069}; \citetalias{Arcodia+2021:eroqpes}; \citealp{Sukova+2021:qpes,Zhao+2021:qpes_star,Xian+2021:qpes_collisions,Metzger+2022:qpes}). Such systems could also emit gravitational wave signals detectable by future detectors like LISA and Tianqin \citep[e.g.,][but see \citealp{Chen+2021:qpebkg}]{Amaro-Seoane+2007:emrilisa,Babak+2017:emris,Zhao+2021:qpes_star}, and represent an electromagnetic counterpart of the so-called extreme mass-ratio inspirals \citep[EMRIs; e.g.,][]{Hils+1995:emris}.
	
	EMRI-related origin scenarios for QPEs are consistent overall with their multi-wavelength observational properties. The presence of EMRIs is, similarly to tidal disruption events (TDEs), only observable for relatively low black hole masses 
	\citep[e.g. $M_{BH}\lessapprox \text{few} \times 10^7\,M_{\astrosun}$,][]{Hills1975:tdes,Wevers+2017:tdemasses,Stone+2020:tderates}, 
	as the stellar-mass orbiting bodies can be otherwise silently swallowed. This mass range is consistent with QPE hosts \citep{Wevers+2022:hosts}, and is also where nuclear star clusters are nearly ubiquitous \citep[e.g.,][for a review]{Neumayer+2020:nsc}, which was confirmed for at least GSN069 (\citetalias{Miniutti+2019:qpe1}; \citealp{Sheng+2021:gsntde}) and RX J1301.9+2747 \citep{Shu+2017:Rx}, and which enhances interactions between the nuclear black hole and stellar-mass objects \citep{Rauch1999:bh_nsc,Babak+2017:emris,Neumayer+2020:nsc}. At least \citep[but see][]{Wevers+2022:hosts} the QPE source RX J1301.9+2747 is a young post-starburst galaxy \citep{Caldwell+1999:RXJ}, a type in which an enhanced rate of TDEs is observed \citep[e.g.,][]{Arcavi+2014:tdes,French+2016:tdes} due to higher stellar density \citep{Stone+2016:tdes,Stone+2018:tdf,Law-Smith+2017:tdes,Hammerstein+2021:tdes} compared to other much more common types of quiescent galaxies. GSN 069 shows UV line-ratios consistent with a past TDE \citep{Sheng+2021:gsntde}, which is also supported by the long-term X-ray emission previous to the QPEs discovery (\citealp{Shu+2018:gsn}; \citetalias{Miniutti+2019:qpe1}). Moreover, recently a new QPE candidate was found in the \emph{XMM-Newton} archives \citep{Chakraborty+2021:qpe5cand}, which also showed previous long-term X-ray emission indicative of a past TDE, further strengthening the EMRI interpretation.
	
	However, even if the macro-scenario might have potentially been pinpointed, there are a lot of differences among the various EMRI-related models proposed \citep[e.g.][]{King2020:gsn069,Sukova+2021:qpes,Metzger+2022:qpes,Zhao+2021:qpes_star,Xian+2021:qpes_collisions} and only qualitative comparisons with the QPEs spectral and timing properties have been made so far. With this work, we aim to improve on this by exploring further the available \emph{XMM-Newton} timing data of eRO-QPE1. We describe the data reduction and analysis performed in Section~\ref{sec:data}, we present our main results in Sections~\ref{sec:multiple_bursts} and~\ref{sec:Edep}, whereas we discuss their implications on our current understanding of QPEs in Section~\ref{sec:disc}.
	
	\begin{figure*}[tb]
		\centering
		\includegraphics[width=17cm]{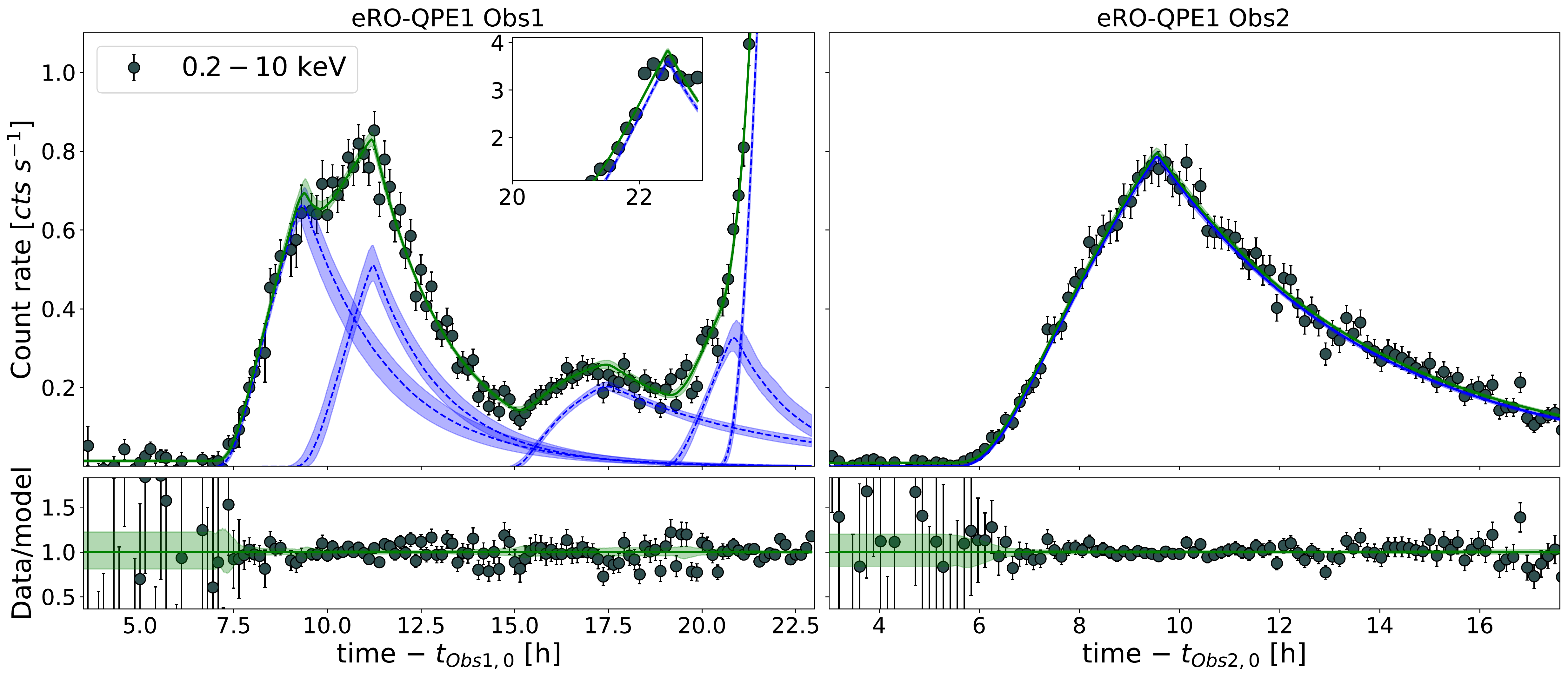}
		\caption{\emph{XMM-Newton} background-subtracted light curve of the two observations of eRO-QPE1, named Obs1 and Obs2, shown in the left and right panels, respectively. The full energy band was used, although we stress that eRO-QPE1 shows little signal above $\gtrsim1.5-2.0\,$keV even in the bright phase \citepalias{Arcodia+2021:eroqpes}. The two observations are separated by roughly a week, the times shown are scaled by the start time of each observation (with modified Julian date $t_{Obs1,0}\sim59057.834$ and $t_{Obs2,0}\sim59065.988$). Top panels show the total model fitted (green line and 1-sigma shaded area), which is composed of one (Obs2) or five (Obs1) individual bursts (blue), each with a profile described by Eq.~\ref{eq:1}. A constant value for the quiescence is also added to the model. The inset in the top left panel shows the peak of the last eruption in Obs1, left out from the linear scale, which was adopted for clearer visualization of the fainter bursts. Bottom panels show the ratio between data and model, where shaded contours show the uncertainty in the model.}
		\label{fig:qpe1_lc}
	\end{figure*}
		
	\section{Data analysis}
	\label{sec:data}
	
\begin{figure}[tb]
		\centering
		\resizebox{\hsize}{!}{
		\includegraphics{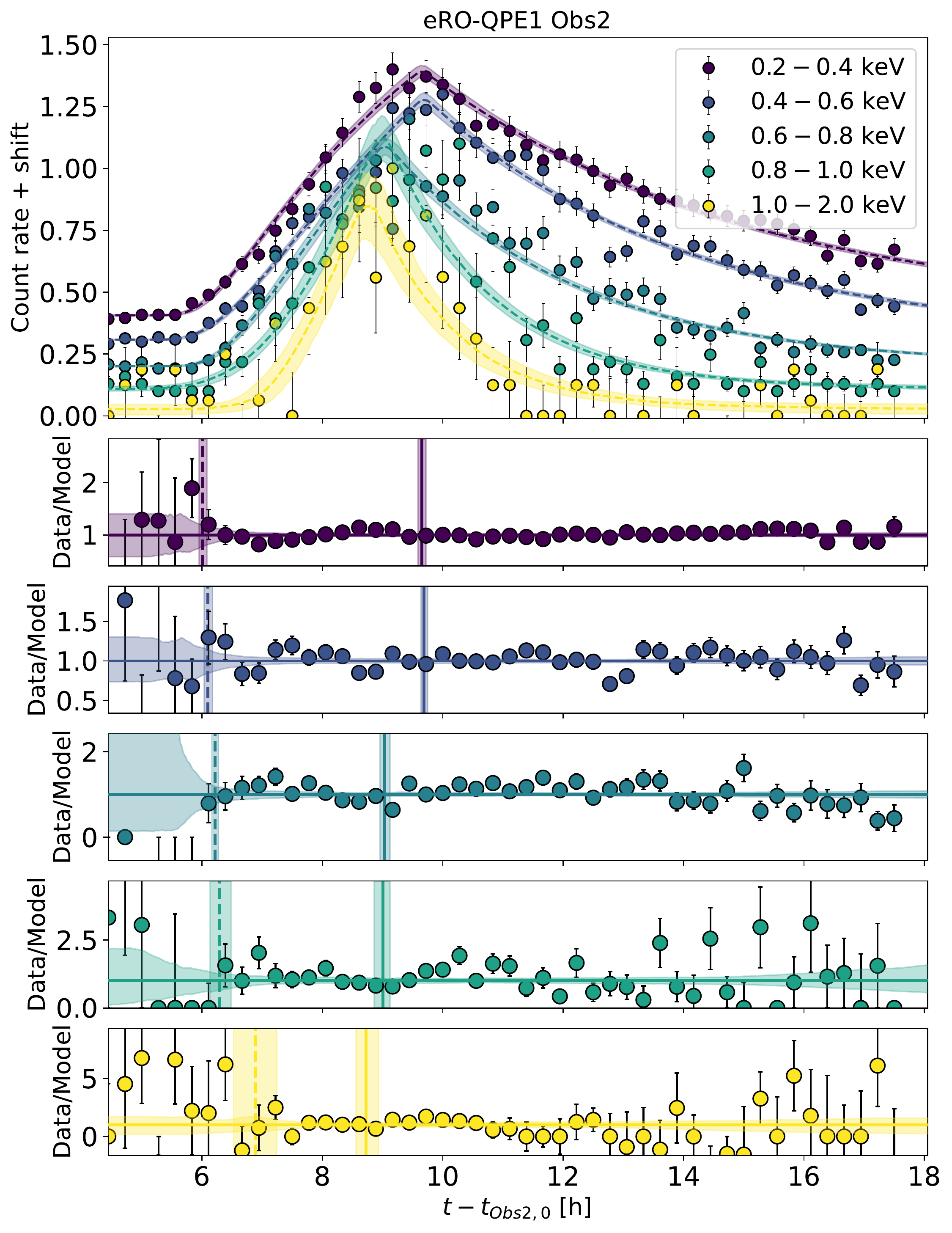}	
		}
		\caption{\emph{XMM-Newton} light curve of Obs2 extracted in small energy bins, represented with different colors as in the legend. Each light curve is vertically shifted with a constant factor for visualization. Lower sub-panels show the data-model residuals in each energy bin, color-coded accordingly; vertical dashed lines (with related 1$\sigma$ uncertainties) indicate when the eruptions start in the given energy bin, while vertical solid lines identify the peak. 
		The model adopted is described by Eq.~\ref{eq:1}, and it is shown in all panels with 1$\sigma$ uncertainty as shaded intervals.}
		\label{fig:qpe1_edep_lc}
	\end{figure}

\begin{figure}[tb]
		\centering
		\includegraphics[width=0.97\columnwidth]{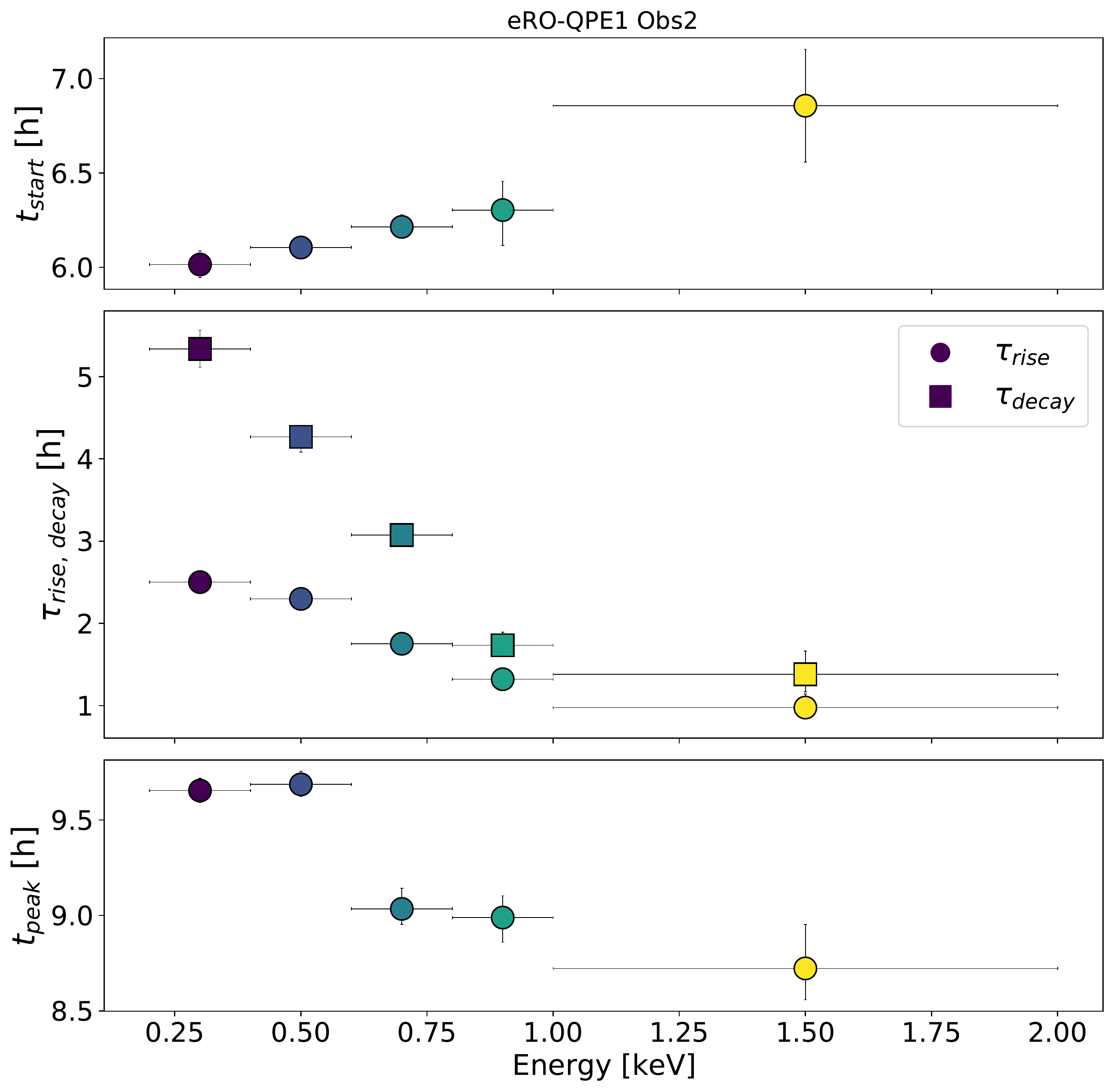}	
		\caption{Energy dependence of QPEs in the observation Obs2 of eRO-QPE1 (related to the light curve shown in Fig.~\ref{fig:qpe1_edep_lc}). We note that $t_{start}$ and $t_{peak}$ are shown in hours elapsed from $t_{Obs2,0}$. QPEs start earlier at lower energies (top panel) and evolve more slowly, in terms of rise and decay (middle panel), and peak at later times (bottom panel) at low energies.}
		\label{fig:qpe1_edep_times}
	\end{figure}

	Two \emph{XMM-Newton} observations of eRO-QPE1, also known as the $z=0.505$ galaxy 2MASS\,02314715-1020112, were analyzed, namely Obs. ID 0861910201 taken on 27 July 2020 (hereafter Obs1) and Obs. ID 0861910301 taken on 4 August 2020 (hereafter Obs2). Details on the processing are described in \citetalias{Arcodia+2021:eroqpes}. We performed timing analysis on the light curve extracted in the full energy range ($0.2-10.0\,$keV), albeit we note that QPEs are very soft events and most of the counts are below $\sim2\,$keV even in the bright phase \citepalias{Miniutti+2019:qpe1,Giustini+2020:qpe2,Arcodia+2021:eroqpes}. We show and discuss results on the QPEs full-band light curve in Fig.~\ref{fig:qpe1_lc} and Section~\ref{sec:multiple_bursts}. We also performed timing analysis divided in small energy bins: $0.2-0.4\,$keV, $0.4-0.6\,$keV, $0.6-0.8\,$keV, $0.8-1.0\,$keV and $1.0-2.0\,$keV, represented throughout the paper with a color map from purple ($0.2-0.4\,$keV) to yellow ($1.0-2.0\,$keV). We show and discuss related results in Figures~\ref{fig:qpe1_edep_lc} to~\ref{fig:qpe1_hr} and Section~\ref{sec:Edep}.

	The modeled profile shape adopted to fit the eruptions was chosen using Obs2, since it shows a single isolated burst (see right panel of Fig.~\ref{fig:qpe1_lc}). It is loosely based on that proposed in \citet{Norris+2005:grb} for gamma-ray bursts, and defined as:
	\begin{equation}
	\begin{cases}
	\label{eq:1}
	A\,\lambda\,e^{\tau_1/(t_{peak} - t_{as} - t)} & \text{if $t<t_{peak}$}\\
	A\,e^{- (t-t_{peak})/\tau_2} & \text{if $t>=t_{peak}$}\\
	\end{cases}       
	\end{equation}
	which is evaluated at zero for times smaller than the asymptote at $t_{peak} - t_{as}$, where $t_{as}=\sqrt{\tau_1\,\tau_2}$. Whereas $A$ is the amplitude at the peak and $\lambda=e^{t_{\lambda}}$ a normalization to join rise and decay, where $t_{\lambda}=\sqrt{\tau_1/\tau_2}$. We note that $\tau_1$ and $\tau_2$ are two characteristic timescales, although only $\tau_2$ is directly related to the decay timescale. Rise and decay times can be defined as a function of $1/e^{n}$ factors with respect to the peak flux, with $n$ being an integer. Following again \citet{Norris+2005:grb} we define rise and decays from the width and asymmetry factors ($w$ and $k$), where $w(n)=\tau_2\,n - \tau_1/(t_{\lambda}\,n) + t_{as}$ and $k(n)= [\tau_2\,n + \tau_1/(t_{\lambda}\,n) - t_{as}]/w$. We define rise and decay timescales as $\tau_{rise}=w\,(1-k)/2$ and $\tau_{decay}=w\,(1+k)/2$ using $n=1$. We also define the start of a burst computing the time at which the flux is $1/e^3$ of the peak value, hence $t_{start}=t_{peak}-w\,(1-k)/2$ with $w$ and $k$ evaluated at $n=3$. We implemented the model and derived posterior probability distributions, Bayesian evidence and Akaike information criterion (AIC) values \citep{Akaike1974:AIC}, using the nested sampling Monte Carlo algorithm MLFriends \citep{Buchner+2014:mlf,Buchner2019:mlf} using the UltraNest\footnote{\url{https://johannesbuchner.github.io/UltraNest/}} package \citep{Buchner2021:ultranest}. In this work, the relative goodness of fit between two models is inferred by comparing AIC values, namely $AIC = 2N_p - 2log\hat{L}$, where $N_p$ is the number of parameters in the model and $log\hat{L}$ is the maximum likelihood of the fit. The fit is considered improved with a more complex model (e.g. with more burst profiles) if its AIC value is smaller, namely if there is a negative difference between the respective AIC values (quantity referred to as $\Delta \text{AIC}_{m2, m1} = \text{AIC}_{m2} - \text{AIC}_{m1}$, where $\text{AIC}_{m2}$ is that of the more complex model). For each model comparison, we also checked the difference in logarithmic Bayesian evidence and find results consistent with the $\Delta \text{AIC}$. We note that other burst profiles, for instance the model as defined in \citet{Norris+2005:grb} - which is smooth at the peak - or a model with a Gaussian rise and exponential decay \citep{vanVelzen+2019:tdeprofile}, were tested but obtained worse fits on Obs2, and were therefore discarded. Throughout the paper, we quote median values of the posterior chains, with related 16th and 84th percentile values, unless otherwise stated.
	
	For the comparison with eRO-QPE1 in Fig.~\ref{fig:qpe2_lc}, we also analyzed and presented \emph{XMM-Newton} data of eRO-QPE2 \citepalias{Arcodia+2021:eroqpes} and GSN\,069 \citepalias{Miniutti+2019:qpe1} and NICER data of eRO-QPE1 \citepalias{Arcodia+2021:eroqpes}. For eRO-QPE2, data processing for Obs. ID 0872390101 (August 2020) was performed and explained in details in \citetalias{Arcodia+2021:eroqpes}. Similarly, details of the processing of NICER (Obs. ID 3201730103) data of eRO-QPE1 are reported in \citetalias{Arcodia+2021:eroqpes}. For GSN\,069, we processed Obs. ID 0831790701 (January 2019) and 0851180401 (May 2019) using standard procedures (SAS v. 18.0.0 and HEAsoft v. 6.25), and the source (background) region was extracted within a circle of $50''$ centered on the source (in a source-free region). Photons were extracted between 0.2-10.0 keV.
	
	\section{The presence of multiple overlapping bursts}
	\label{sec:multiple_bursts}
	
	Although not exactly periodic, eruptions observed in a given QPE-emitting source have so far been estimated as single isolated bursts with a somewhat regular recurrence pattern \citepalias{Miniutti+2019:qpe1,Giustini+2020:qpe2,Arcodia+2021:eroqpes}. For instance, eruptions in GSN\,069 \citepalias{Miniutti+2019:qpe1} and eRO-QPE2 \citepalias{Arcodia+2021:eroqpes} show a repeating alternation of longer and shorter recurrence times after alternating stronger and weaker bursts, respectively (\citetalias{Miniutti+2019:qpe1}; \citealp{Xian+2021:qpes_collisions}; our Fig.~\ref{fig:qpe2_lc}). 

	In Fig.~\ref{fig:qpe2_lc} we show the light curves of both eRO-QPE2 and GSN\,069 (top and middle panel, respectively) divided in strong-weak-strong QPE-cycles, each cycle defined as the separation between two strong QPEs and overlapped in time, by scaling every data point by the $t_{peak}$ of the first eruption of each cycle. Dashed vertical lines show the time corresponding to half of the separation between two strong bursts (i.e. between the bursts aligned to zero in Fig.~\ref{fig:qpe2_lc}), highlighting that weak bursts occur systematically later in time than half of the strong-strong recurrence. Furthermore, all superimposed strong (weak) eruptions appear to be overall compatible with one another.
	
	In general, eRO-QPE1 immediately stands out with respect to the other known QPEs \citep[e.g., see][]{Chakraborty+2021:qpe5cand}, due to its peak X-ray luminosity ($\gtrsim10^{43}\,$erg\,s$^{-1}$) and evolving timescales (mean recurrence of $\sim0.8$\,d), which are larger by an order of magnitude \citepalias{Arcodia+2021:eroqpes}. Contrary to eRO-QPE2 and GSN\,069, a quick look at a similar visualization for eRO-QPE1 (bottom panel of Fig.~\ref{fig:qpe2_lc}) also argues for a more complex evolution and shape of the eruptions. To investigate the latter, we show in Fig.~\ref{fig:qpe1_lc} the two deep \emph{XMM-Newton} observations of eRO-QPE1, Obs1 (left) and Obs2 (right), with the related best fit model and data/model ratios. It is evident from the data points alone that the Obs2 (right panel) shows evidence of a single burst, similar to what is observed for eRO-QPE2 and GSN\,069 (in Fig.~\ref{fig:qpe2_lc}), albeit on much longer timescales and with an enhanced asymmetry. Instead, Obs1 (left panel) clearly shows a much more complex event. We first fit to the much simpler Obs2 the model in Eq.~\ref{eq:1} with a constant plateau in addition, obtaining a very good fit and residuals throughout rise, peak and decay. This flare template was adopted as a reference shape, which we then applied to Obs1, with the assumption that we could model its more complex behavior as a linear combination of the simple template. We started with a sum of three different bursts and increased their number by comparing the related AIC values and by visualizing the fit residuals. A model with four bursts improves the fit with a $\Delta\text{AIC}\sim-167.8$. Residuals around $t-t_{Obs1,0}\sim20$\,h (see Fig.~\ref{fig:qpe1_lc_xmm1_3and4}) were significant improved by adding a fifth burst, which improved the four-bursts model with $\Delta\text{AIC}\sim-91.8$. Therefore, we obtained that five independent profiles are needed (see Appendix~\ref{sec:xmm1} for more details). Since the model adopted (Eq.~\ref{eq:1}) fits the simpler Obs2 well, we consider it unlikely that Obs1 needs a completely different and more complicated shape for the eruptions, but rather a combination of the simple burst template used for the simpler Obs2. A further relevant question is whether the complex behavior seen in Obs1 was a one-time (or rare) event, or whether it recurs throughout the long-term evolution of eRO-QPE1. We must note that studying the latter with \emph{XMM-Newton} is observationally expensive, in fact this will be the focus of an upcoming work (Arcodia et al., in prep.).

	As a sanity check, we tested whether any statistical evidence of more complex structure is needed for Obs2 of eRO-QPE1. We obtained that Obs2 is best described with one single eruption (see Appendix~\ref{sec:xmm1} and Fig.~\ref{fig:qpe1_lc_xmm2_2bursts} for more details), which validates the use of its template to fit Obs1. Moreover, we performed a similar test on the seemingly much simpler QPE source eRO-QPE2 (see Appendix~\ref{sec:xmm1}). If a small degree of asymmetry is accounted for, no superposition of individual bursts is required to model the eruptions in eRO-QPE2 (see Appendix~\ref{sec:xmm1} and Fig.~\ref{fig:qpe2_tests}). Therefore no significant substructures are present in Obs2 of eRO-QPE1, nor in the bursts of eRO-QPE2.
	
	\section{The energy dependence of QPEs}
	\label{sec:Edep}

	Since the discovery of QPEs, it was quickly noticed that once the light curve is decomposed into small energy bins, eruptions appear to peak at later times at lower energies and are broader compared to higher energies \citepalias{Miniutti+2019:qpe1,Giustini+2020:qpe2}. However, it was not investigated before in QPEs observations whether there is an energy dependence at the start of the rise of QPEs. This is mostly due to the relatively lower signal-to-noise ratio in the energy-resolved light curves and shorter evolution timescales in the first QPE sources \citepalias{Miniutti+2019:qpe1,Giustini+2020:qpe2}. Finding or excluding an energy dependence of the QPEs start is however important to understand whether QPEs are consistent with an emission component with a given spectrum simply getting brighter in luminosity, or if instead the spectrum evolves in temperature/energy over time during the start, as it seems to do during the rise \citepalias{Miniutti+2019:qpe1}. In fact, some sort of compact accretion flow should be present in QPEs, as they show a stable spectrum during quiescence which is as bright as $\approx0.3-1.6\times10^{41}$\,erg\,s$^{-1}$ in the soft X-ray band (\citetalias{Miniutti+2019:qpe1,Giustini+2020:qpe2,Arcodia+2021:eroqpes}; \citealp{Chakraborty+2021:qpe5cand}). Furthermore, eRO-QPE1 is ideal to test the hypothesis of an energy dependence of the QPEs start times, thanks to its luminosity and timescales which are one order of magnitude larger than all the other QPEs found so far (\citetalias{Miniutti+2019:qpe1,Giustini+2020:qpe2,Arcodia+2021:eroqpes}; \citealp{Chakraborty+2021:qpe5cand}).
	
	\begin{figure}[!htb]
		\centering
		\includegraphics[width=0.93\columnwidth]{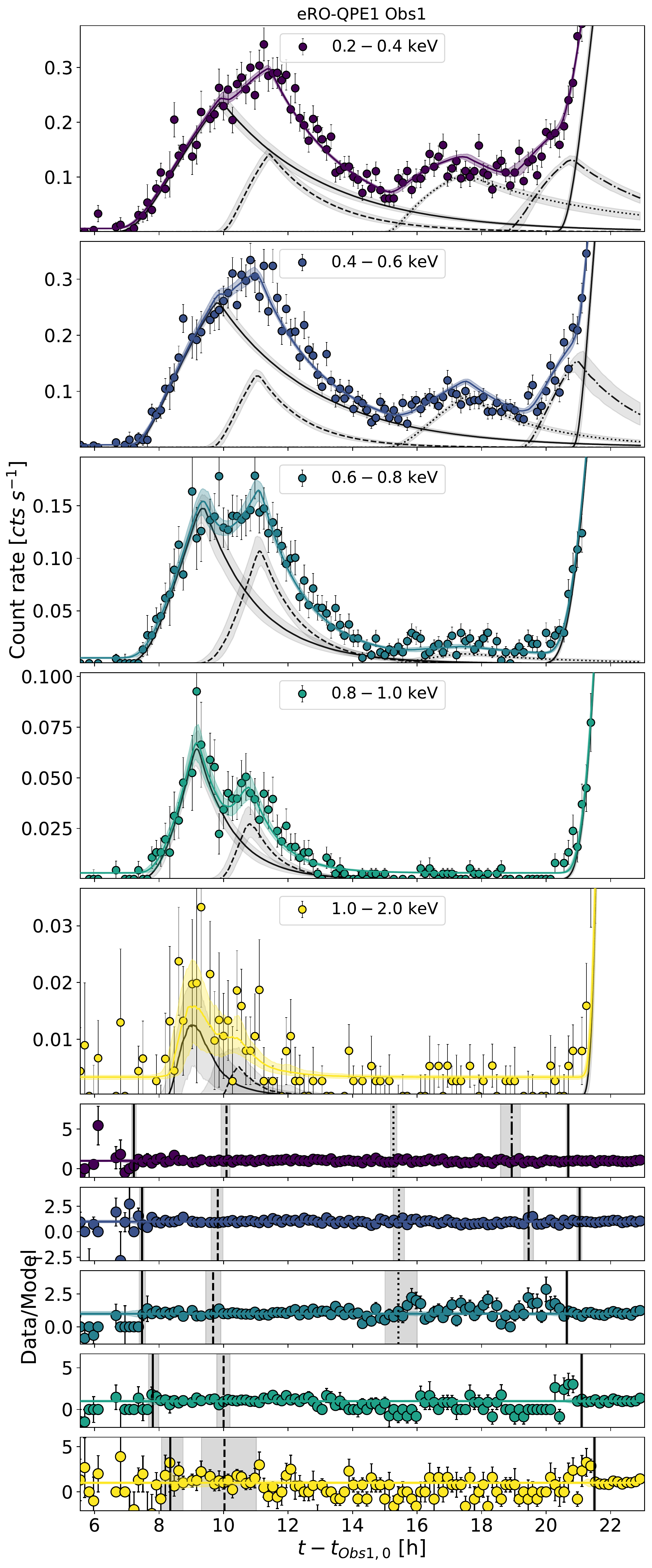}	
		\caption{Same as Fig.~\ref{fig:qpe1_edep_lc}, but for Obs1. Individual burst profiles are shown in black, with different line styles. Five bursts are required for the $0.2-0.4\,$keV and $0.4-0.6\,$keV bins, while four for $0.6-0.8\,$keV and three in the remaining energy bins (see Table~\ref{tab:logz_obs1_Edep}). 
		The lower panels show data/model ratios. Vertical lines indicate $t_{start}$ of each burst at all energies, with the line style matching the one used for the burst profiles in the upper panels with the light curves. Shaded intervals show $1\sigma$ uncertainties.}
		\label{fig:qpe1_edep_lc_xmm1}
	\end{figure}

	\begin{figure}[tb]
		\centering
		\includegraphics[width=0.99\columnwidth]{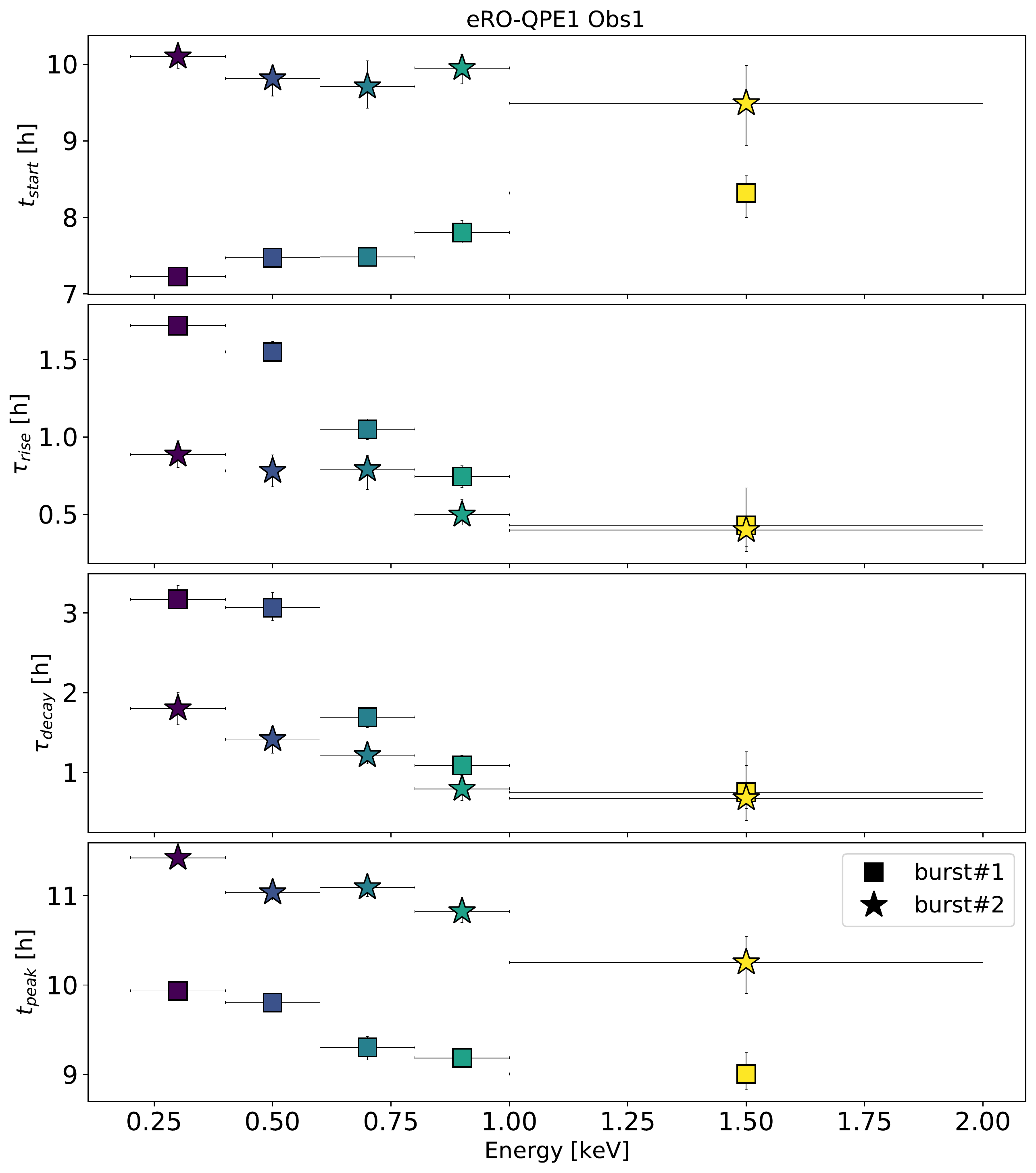}	
		\caption{Same as Fig.~\ref{fig:qpe1_edep_times}, but for the first two bursts of Obs1 (e.g. see Fig.~\ref{fig:qpe1_edep_lc_xmm1}). Squares are used in all panels for the first burst, stars for the second. The energy dependence of $\tau_{rise}$, $\tau_{decay}$ and $t_{peak}$ (from second-to-top to bottom) is consistent with Obs1 and other QPEs (\citetalias{Miniutti+2019:qpe1,Giustini+2020:qpe2}; \citealp{Chakraborty+2021:qpe5cand}). 
		}
		\label{fig:qpe1xmm1_edep_times}
	\end{figure}
	
	Hence, we extracted light curves in small energy bins for both Obs1 and Obs2 (Section~\ref{sec:data}). We start from Obs2 due to its simplicity. Fig.~\ref{fig:qpe1_edep_lc} shows Obs2 divided in energy bins, with the related best-fit models obtained using Eq.~\ref{eq:1} and data/model ratios in the lower panels. Shaded intervals indicate $1\sigma$ uncertainties on the model. We also highlight the median value (with shaded 1$\sigma$ uncertainties) of the fit posteriors of $t_{start}$ and $t_{peak}$ (see Section~\ref{sec:data}) with a vertical dashed and solid line, respectively. Moreover, we show in Fig.~\ref{fig:qpe1_edep_times} $t_{start}$, $\tau_{rise}$, $\tau_{decay}$ and $t_{peak}$ as a function of energy, as obtained from our light curve fit. The overall picture is that we confirm the previously observed trend for rise, peak and decay times, namely that QPEs evolve faster and peak earlier at higher energies \citepalias{Miniutti+2019:qpe1,Giustini+2020:qpe2}. Furthermore, for the first time we can resolve the start of the eruptions and find that QPEs start earlier at lower energies (see top panel of Fig.~\ref{fig:qpe1_edep_times}).

	We show in Fig.~\ref{fig:qpe1_edep_lc_xmm1} the analogous of Fig.~\ref{fig:qpe1_edep_lc}, but for Obs1, confirming the trend found in Obs2 even during a more complex duty cycle. We fit all light curves in small energy bins (see Section~\ref{sec:data}) with a model including three, four and five bursts. We recursively selected the best fit model in each energy bin by comparing their AIC values. We report the corresponding $\Delta \text{AIC}$ values in Table~\ref{tab:logz_obs1_Edep}. Intriguingly, we obtained that burst number three and four of the full-energy light curve (see Fig.~\ref{fig:qpe1_lc} left) not only have the lowest amplitude, but are also much colder than the brighter eruptions. In particular, little or no significant signal is detected above $\sim0.6-0.8\,$keV (Fig.~\ref{fig:qpe1_edep_lc_xmm1}). Therefore, good enough fits are obtained with four bursts in the $0.6-0.8\,$keV bin and three at the highest energy bins (e.g. $\gtrsim0.8\,$keV). The lower panels of Fig.~\ref{fig:qpe1_edep_lc_xmm1} show the data/model ratios, while vertical lines show the fit $t_{start}$ for each burst (with line style matching the one used in the upper panels for the burst profile). For bursts starting with some overlap with the previous, $t_{start}$ is obviously less constrained. However, $t_{start}$ in the first and the last (which is also the brightest) burst of Obs1 does show an increase with energy, consistently with Obs2. We also note that in the energy bins between $0.6-1.0\,$keV the difference between the two peaks around $t-t_{Obs1,0}\sim10\,$h is enhanced (Fig.~\ref{fig:qpe1_edep_lc_xmm1}), confirming the need of a combination of burst profiles rather than a more complex single burst model. We show in Fig.~\ref{fig:qpe1xmm1_edep_times} the fit $t_{start}$, $\tau_{rise}$, $\tau_{decay}$ and $t_{peak}$ of these two bursts in Obs1 as a function of energy (shown with different symbols as in the legend). The energy dependence of $\tau_{rise}$, $\tau_{decay}$ and $t_{peak}$ (from second-to-top to bottom) is consistent with Obs1 and other QPEs (\citetalias{Miniutti+2019:qpe1,Giustini+2020:qpe2}; \citealp{Chakraborty+2021:qpe5cand}). $t_{start}$ of the first burst is also consistent with Obs2 (Fig.~\ref{fig:qpe1_edep_times}), in that it increases with energy. Instead, in the second burst $t_{start}$ seems to be compatible within uncertainties across energies. Given the likely degeneracy with the decay of the previous burst, we refrain from over-interpreting this dependency.		
	
	Hence, from both the simpler Obs2 and the more complex Obs1, we inferred an earlier $t_{start}$ of the eruptions at lower energies. We stress that this is unlikely to be a model-dependent artifact as it is clearly observable by eye in the data points of Fig.~\ref{fig:qpe1_edep_lc} (e.g. comparing the lowest and highest energy bins). Finally, this result can not be a spurious effect due to the instrument sensitivity, as \emph{XMM-Newton} is much more sensitive around $\sim1\,$keV than at $\sim0.3\,$keV; and a possible incorrect determination of the counts redistribution in the current \emph{XMM-Newton} calibration\footnote{\href{https://xmmweb.esac.esa.int/docs/documents/CAL-TN-0018.pdf}{https://xmmweb.esac.esa.int/docs/documents/CAL-TN-0018.pdf}} would be relevant for the opposite case, namely if higher energy photons were detected first.
	
	Finally, we note that in other QPE sources an energy dependence of the bursts amplitude is observed (\citetalias{Miniutti+2019:qpe1,Giustini+2020:qpe2}; \citealp{Chakraborty+2021:qpe5cand}). It is usually defined with respect to a well-detected and unabsorbed quiescence emission and leads to larger amplitudes in higher energy bins. Instead, in eRO-QPE1 the quiescence is not robustly detected and the amplitude would be computed against the background spectrum of \emph{XMM-Newton}, therefore it would lead to a misleading comparison. We propose in the next Section an alternative way to compare the spectral evolution across different QPE sources.
	
	\begin{figure*}[t]
		\centering
		\includegraphics[width=17cm]{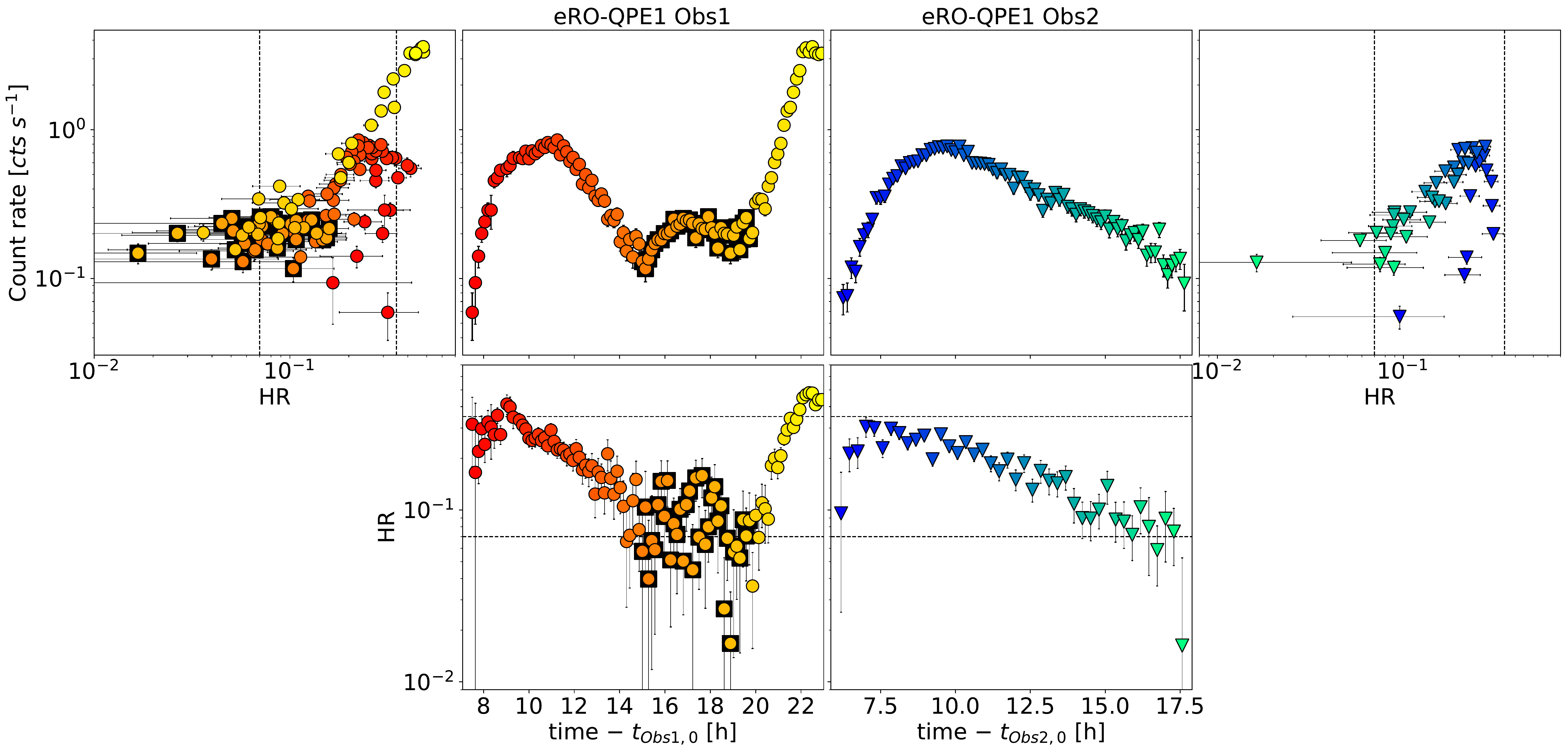}	
		\caption{The left (right) half of the plot shows results for Obs1 (Obs2) of eRO-QPE1, represented with colored circles (triangles), for which darker to lighter colors with a red-to-yellow (blue-to-green) color map indicate the increase of time in the observation. \emph{Top central panels}: the count rate (CR) light curve as in Fig.~\ref{fig:qpe1_lc}, but here in logarithmic units and excluding the quiescence. \emph{Bottom panels}: the evolution of hardness ratio ($\text{HR} = H/(H+S) =  \text{CR}_{0.6-2.0\,keV}/\text{CR}_{0.2-2.0\,keV}$) over time.  \emph{External panels}: the HR versus CR plot. In the panels related to Obs1, we highlight the central data points with black squares to ease the comparison across panels. We show the range spanned by Obs2 in HR with black dashed lines across different panels. The HR reached at the peak of the eruptions is count rate dependent and eruptions undergo an anti-clockwise hysteresis cycle in the HR versus CR plot (e.g. top right).}
		\label{fig:qpe1_hr}
	\end{figure*}
	
	\begin{figure}[t]
		\centering
		\includegraphics[width=0.7\columnwidth]{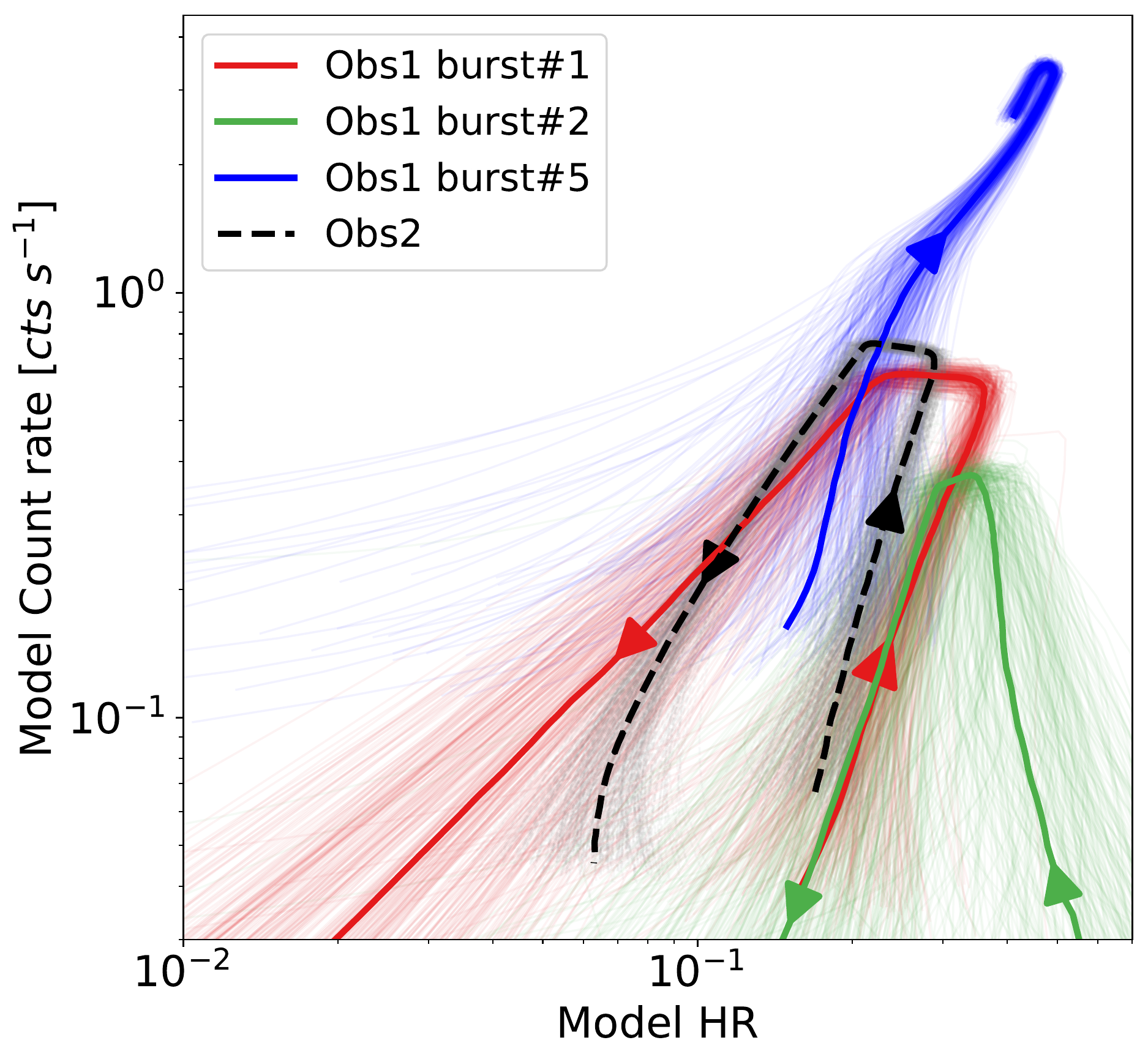}	
		\caption{Hardness ratio (HR) versus count rate (CR) computed from the modeled profiles of the bursts in Obs1 (color coded as in the legend) and Obs2 (dashed black line). For a visual representation of the model profiles refer to Fig.~\ref{fig:qpe1_lc}. For Obs1, we have used the profiles of bursts which are clearly detected in all the energy bands used for the definition of HR. The thick solid lines represent the median of the modeled tracks, while the thinner lines highlight the span within the model posteriors. An anti-clockwise hysteresis (highlighted with arrows on the tracks) is found for all modeled eruptions, which is also what pure data indicate (Fig.~\ref{fig:qpe1_hr}).}
		\label{fig:qpe1_hr_model}
	\end{figure}
	
	\section{Hardness ratio of the eruptions}
	\label{sec:hr}
	
	We further tested the energy dependence of QPEs by computing the hardness ratio of Obs1 and Obs2, which confirms the results above in a model-independent way. It was computed as $\text{HR} = H/(H+S) = \text{CR}_{0.6-2.0\,keV}/\text{CR}_{0.2-2.0\,keV}$, where CR is the count rate in the related energy bands. We report in the left (right) half of Fig.~\ref{fig:qpe1_hr} results for Obs1 (Obs2), represented with circles (triangles), for which darker to lighter colors indicate the increase of time: the CR light curve is shown in the top middle panel, while the HR evolution over time at the bottom and the HR versus CR plot in the top external panel.
	
	For simplicity, let us start with Obs2. The evolution of HR over time indicates a gradual spectral hardening during the rise and softening during decay, as expected from previous results (\citetalias{Miniutti+2019:qpe1,Giustini+2020:qpe2,Arcodia+2021:eroqpes}; \citealp{Chakraborty+2021:qpe5cand}). However, the spectrum does not harden throughout the whole rise, in fact HR remains constant and then slightly decreases during the second part of the CR rise. We interpret this as the effect of the energy dependence of QPEs, as eruptions are known to peak later in time at lower energies (as we have just shown in Fig.~\ref{fig:qpe1_edep_lc} to~\ref{fig:qpe1xmm1_edep_times} and, e.g., \citetalias{Miniutti+2019:qpe1,Giustini+2020:qpe2}; \citealp{Chakraborty+2021:qpe5cand}), therefore enhancing the softening. Notably, it is the first time that QPEs are shown to undergo an anti-clockwise hysteresis in an HR versus CR plot (top right for Obs2), namely there is not a unique spectral shape for a given count rate during both rise and decay. As a matter of fact, the source is softer during decay with respect to the rise for a compatible total count rate. Therefore, QPEs are not only asymmetric in their timing profiles, but also spectrally.
	
	From a comparison with Obs1 (left half of Fig.~\ref{fig:qpe1_hr}), we notice that the maximum hardness reached by an eruption increases with its peak count rate (bottom left panel and top left panels). The first part of Obs1 (redder colors) seems to show an anti-clockwise hysteresis similarly to the eruption in Obs2 and the same is apparent for the final section of Obs1 (yellower colors), although the observation stopped just before the decay. Nonetheless, from the end of both the CR light curve and the HR evolution of Obs1 (top central-left and bottom left panel of Fig.~\ref{fig:qpe1_hr}), a flattening and possible start of decrease can be seen. Therefore, this suggests that the brightest eruption started a compatible anti-clockwise hysteresis, albeit shifted to higher total count rates. Data points related to the weaker state in the middle of Obs1 (highlighted with black squares) seem to confirm the exceptional softness and weakness of the source (though still brighter than the quiescence), as hinted in Fig.~\ref{fig:qpe1_edep_lc_xmm1} as well. We also note that the substructures in the first part of Obs1 are less remarkable in Fig.~\ref{fig:qpe1_hr}, although they remain quite evident in Fig.~\ref{fig:qpe1_edep_lc_xmm1}. As a sanity check for our model-dependent analysis (Sections~\ref{sec:multiple_bursts} and~\ref{sec:Edep}), we compared in Fig.~\ref{fig:qpe1_hr_model} the HR vs CR of the modeled profiles of the single eruptions of Obs1 (see Fig.~\ref{fig:qpe1_edep_lc_xmm1}; albeit only for the bursts clearly detected in all energy bands used for the definition of HR) with that of Obs2 (black dashed line). For the latter, the comparison with the plot showing pure data (top right of Fig.~\ref{fig:qpe1_hr}) is straightforward. Intriguingly, a clear anti-clockwise hysteresis is also evident for each of the bursts in Obs1 separately. This indicates the robustness of our decomposition of Obs1 in multiple bursts, as we managed to retrieve for them a spectral behavior compatible with that of the isolated burst of Obs2. 
	
	\section{Discussion}
	\label{sec:disc}
	\subsection{Two populations of QPEs?}	
	\label{sec:disc_2classes}
		
	A first look at QPEs observations had so far suggested that in a given source, eruptions would occur more or less regularly, modulo some scatter in the quasi-period, with single seemingly isolated bursts \citepalias{Miniutti+2019:qpe1,Giustini+2020:qpe2,Arcodia+2021:eroqpes}. This seems to be indeed the case for at least two of the four secure QPEs, namely GSN 069 \citepalias{Miniutti+2019:qpe1} and eRO-QPE2 \citepalias{Arcodia+2021:eroqpes}. In these two sources, eruptions occur with an alternation between long and short recurrence times, happening after relatively stronger and weaker bursts, respectively (see Fig.~\ref{fig:qpe2_lc}; \citetalias{Miniutti+2019:qpe1}; \citealp{Xian+2021:qpes_collisions}). Moreover, strong (weak) eruptions have so far appeared to be overall comparable in their profiles with other strong (weak) ones, lacking any significant and observable signature of overlapping bursts (Fig.~\ref{fig:qpe2_lc}, Appendix~\ref{sec:xmm1}). Instead, in this work we have shown that in the source eRO-QPE1, its timing properties can be far more complex and irregular. In particular, we have observed the presence of cycles with single isolated bursts (Fig.~\ref{fig:qpe1_lc} right, i.e. similar to what we expected based on other QPEs), as well as other cycles in which we observed multiple overlapping bursts with different amplitudes (Fig.~\ref{fig:qpe1_lc} left). This complex timing behavior may also be present in RX J1301.9+2747 \citepalias{Giustini+2020:qpe2}, though to a less dramatic extent, with the presence of two bursts with much lower amplitude than all the others (Giustini et al., in prep.). This might hint that two sub-classes of QPEs exist, the first sub-group being GSN 069 \citepalias{Miniutti+2019:qpe1} and eRO-QPE2 \citepalias{Arcodia+2021:eroqpes}, while the second populated by RX J1301.9+2747 \citepalias{Giustini+2020:qpe2} and eRO-QPE1 \citepalias{Arcodia+2021:eroqpes}. Although only by finding more we can infer whether this is the case, or whether instead QPEs timing properties are a continuum which spans from rather simple and regular, as GSN\,069 and eRO-QPE2, to more complex and irregular, as eRO-QPE1 passing through RX J1301.9+2747.
	
	Given the many similarities in the energy-dependent properties of all QPEs (\citetalias{Miniutti+2019:qpe1,Giustini+2020:qpe2,Arcodia+2021:eroqpes}; \citealp{Chakraborty+2021:qpe5cand}), we can consider the evidence, here reported (Section~\ref{sec:Edep}, Fig.~\ref{fig:qpe1_edep_lc} and~\ref{fig:qpe1_edep_times}), of QPEs starting earlier at lower energies in eRO-QPE1, to be valid for all QPEs. Conversely, more complex timing behavior was so far only observed to happen for eRO-QPE1 and RX J1301.9+2747 (to a less dramatic extent; Giustini et al., in prep.). Based on the current theoretical knowledge, it is possible that there are two different physical mechanisms in place for the two putative QPE sub-groups. However, given the many common observed properties within all known QPEs, we can speculate that there could be one main mechanism, be it star-disk collisions\footnote{We will use this term also referring to collisions between a disk and a stellar-mass compact object.} \citep[e.g.,][]{Sukova+2021:qpes,Xian+2021:qpes_collisions}, Roche-lobe overflow (RLOF) from one or two stars \citep[e.g.,][]{Zhao+2021:qpes_star,Metzger+2022:qpes} or an alternative model even yet to be proposed.
	
	\subsection{Implications for the origin of QPEs}	
	
	Based on our findings, any model that aims to explain QPEs must be able to produce both regularly spaced eruptions - which may appear with an alternation of shorter and longer cycles - in a given source or, alternatively, a more erratic behavior with less clear periodicity and different amplitudes (during some of the cycles at least, e.g. Fig.~\ref{fig:qpe1_lc}). Even if an in depth modeling of our new results is beyond the scope of this work, we speculate here their impact on models which involve star-disk collisions \citep[e.g.,][]{Xian+2021:qpes_collisions} and RLOF from one or more stellar companions \citep[e.g.,][]{Zhao+2021:qpes_star,Metzger+2022:qpes}, which were proposed for QPEs. In fact, current models of a pure lensing scenario \citep[e.g.,][]{Ingram+2021:qpes_lensing} seem now disfavored by the known energy dependence of QPEs, the exact shape and amplitude of the bursts \citep{Ingram+2021:qpes_lensing} and also considering our new results (Section~\ref{sec:multiple_bursts}), which would need a very complex time-varying gravitational geometry.

	In the framework of QPEs from star-disk collisions \citep[e.g.,][]{Sukova+2021:qpes,Xian+2021:qpes_collisions}, a burst is produced at every crossing, each of which is assumed to leave the companion mostly unaffected over the time span studied \citep{Dai+2010:stardisk,Pihajoki2016:bhimpacts}, although long-term perturbations of the orbit are relevant in the case of a normal star \citep[e.g.,][]{Syer+1991:stardisk,Vokrouhlicky+1993:stardisk,MacLeod+2020:stardisk}. One possible outcome of these collisions is the production of a shock at the impact region within the disk \citep{Pihajoki2016:bhimpacts,Sukova+2021:qpes}. These in turns produce two fountains of expanding gas, which eventually radiate as they become optically thin \citep{Lehto+1996:oj287,Ivanov+1998bhstar,Pihajoki2016:bhimpacts}, or, alternatively, episodic ejections of blobs from the accretion flow \citep{Sukova+2021:qpes}. Only generic predictions have been made for the expected spectral and timing evolution of the resulting emission component \citep[e.g.,][]{Semerak+1999:bhstar,Dai+2010:stardisk}, although it has been suggested that Bremsstrahlung or thermal radiation would be expected in the fountains scenario \citep[e.g.,][]{Lehto+1996:oj287,Pihajoki2016:bhimpacts}, which are both an overall good description for the QPEs spectra \citepalias{Miniutti+2019:qpe1,Giustini+2020:qpe2,Arcodia+2021:eroqpes}. We note that the QPEs' energy dependence (e.g. Fig.~\ref{fig:qpe1_edep_lc} and~\ref{fig:qpe1_edep_times}) could be qualitatively reproduced if the collisions induce heating fronts propagating in the disk \citep[e.g.,][]{Meyer1984:transwaves}, although these shock waves are shown to dramatically destroy the inner disk \citep{Chan+2019:tdeagn}, which is in tension with the stable quiescence detected in QPEs right after the eruptions \citepalias{Miniutti+2019:qpe1,Giustini+2020:qpe2,Arcodia+2021:eroqpes}. Moreover, such a model inherently assumes that throughout several collisions the system is in roughly the same state before each interaction (e.g. intended as the two collisions per orbit), in terms of physical conditions in both the the accretion flow and the star \citep[e.g.,][]{Ivanov+1998bhstar,Dai+2010:stardisk,Pihajoki2016:bhimpacts}. This is in tension with our results on eRO-QPE1, which shows cycles with single isolated bursts sometimes (Fig.~\ref{fig:qpe1_lc} right) and more complex behavior other times (left panel). This is particularly puzzling since the two observations were taken one week apart and, assuming a periodicity of the order of less than a day \citepalias{Arcodia+2021:eroqpes}, just a handful of cycles have occurred in between. Therefore, despite being a promising scenario \citep[e.g., see][for GSN 069]{Xian+2021:qpes_collisions}, more quantitative predictions from these models are needed to conclusively state whether they fail in reproducing the more complex timing observations of eRO-QPE1 (e.g. Fig.~\ref{fig:qpe1_lc}) and RX J1301.9+2747.
	
	In the model proposed by \citet{Metzger+2022:qpes}, two stellar EMRIs occur in a co-planar orbit around the central black hole, and the observed QPEs periodicity implies that they are counter-rotating. At least one star undergoes RLOF at each flyby and mass is accreted towards the black hole interacting with - and replenishing - an accretion disk, the presence of which is supported by the stable quiescence usually observed in between QPEs \citepalias{Miniutti+2019:qpe1,Giustini+2020:qpe2,Arcodia+2021:eroqpes}. Possible differences in the mass transferred to the black hole can occur from cycle to cycle due to changes in the separation between the two stars, influenced by residual eccentricity in the system \citep{Metzger+2022:qpes}. This could indeed provide the required diversity in the eruptions. In addition, it is possible that during a given cycle, or most of the cycles, a single star undergoes RLOF, but occasionally both stars do, with a slightly delayed mass transfer rate. As much as this is an extremely complex interaction to model, it is qualitatively in accord with our observations of eRO-QPE1, which show isolated bursts (e.g. Fig.~\ref{fig:qpe1_lc} right) as well as more complex cycles (e.g. Fig.~\ref{fig:qpe1_lc} left). We speculate that this is also a promising model for RX J1301.9+2747 \citepalias{Giustini+2020:qpe2}. Interestingly, if one of the two EMRIs is a star and the second is a compact object, only the star can undergo RLOF and this scenario is worth exploring for GSN 069 and eRO-QPE2, which do not show evidence of multiple overlapping bursts. Other models with a single EMRI \citep{King2020:gsn069,Zhao+2021:qpes_star} would appear similar to the latter double-EMRI model with one star and one compact object, whilst they would be harder to reconcile with eRO-QPE1 and RX J1301.9+2747. Moreover, \citet{Metzger+2022:qpes} show that the post-peak energy dependence of QPEs is reproduced using one-dimensional spreading disk equations \citep[e.g., see][]{Pringle1981:accr} to model the evolution of the matter flowing in towards the black hole. However, more dedicated models or simulations are needed to compare with Fig.~\ref{fig:qpe1_edep_lc}-\ref{fig:qpe1xmm1_edep_times} and model the early phases, when most likely any steady-state assumption for the accretion flow is violated.
	
	\section{Conclusions}
	
	QPEs are the new frontier of variable accretion onto massive black holes (\citetalias{Miniutti+2019:qpe1,Giustini+2020:qpe2,Arcodia+2021:eroqpes}; \citealp{Chakraborty+2021:qpe5cand}). They are able to light up the nuclei of otherwise faint and unnoticed low-mass galaxies \citepalias{Arcodia+2021:eroqpes}, therefore providing a new channel to activate transients accretion events onto the black holes at their center. However, these peculiar repeating X-ray blasts still remain unexplained. Lately, models involving one or two stellar-mass companions around the central black hole have gathered significant attention (\citealp{King2020:gsn069}; \citetalias{Arcodia+2021:eroqpes}; \citealp{Sukova+2021:qpes,Metzger+2022:qpes,Zhao+2021:qpes_star,Xian+2021:qpes_collisions}). If this is indeed the correct origin scenario, QPEs could also emit low-frequency gravitational waves as EMRIs \citep[e.g.,][but see \citealp{Chen+2021:qpebkg}]{Zhao+2021:qpes_star} and could therefore revolutionize the future of multi-messenger astronomy with \emph{Athena} \citep{Nandra+2013:athena} and \emph{LISA} \citep{Amaro-Seoane+2017:LISA}. 
	
	We here report two new observational results found by taking a closer look at the QPEs in eRO-QPE1 \citepalias{Arcodia+2021:eroqpes}:
	\begin{itemize}
		\item[i)] At times, eruptions in eRO-QPE1 occur as single isolated bursts (Fig.~\ref{fig:qpe1_lc} right), while they can also manifest as a complex mixture of multiple overlapping bursts with very diverse amplitudes (Fig.~\ref{fig:qpe1_lc} left); this is in contrast to other known QPEs, namely GSN\,069 and eRO-QPE2, for which so far only evidence of somewhat orderly bursts with a more regular recurrence pattern was found (Fig.~\ref{fig:qpe2_lc}).\\
		\item[ii)] Studying eruptions in small energy bins, we confirm previous works that found QPEs to peak at later times and to be broader at lower energies (\citetalias{Miniutti+2019:qpe1,Giustini+2020:qpe2}; \citealp{Chakraborty+2021:qpe5cand}), while for the first time we find that QPEs start earlier in time at lower energies (Fig.~\ref{fig:qpe1_edep_lc}-\ref{fig:qpe1xmm1_edep_times}). Furthermore, eruptions appear to undergo an anti-clockwise hysteresis cycle in a plane with hardness ratio versus total count rate, implying that the decay of each eruption is softer than its rise at a compatible total count rate (Fig.~\ref{fig:qpe1_hr} and~\ref{fig:qpe1_hr_model}).
	\end{itemize}

	The first result implies that, if we demand a single trigger mechanism for all QPEs, this should be able to produce both behaviors, regular and complex, in some sources (e.g. in eRO-QPE1 and RX J1301.9+2747), while in others only the more regularly-spaced eruptions (e.g. GSN 069 and eRO-QPE2). The second result implies that the X-ray emitting component is not achromatic, namely the resulting spectrum does not simply brighten and fade in luminosity and it has instead a specific energy dependence over time. If there is indeed an accretion flow around these black holes (with any kind of radiative efficiency and thickness) emitting the quiescence signal, the inferred energy dependence might imply the presence of inward radial propagation during the QPEs rise. Current and future models proposed for QPEs should reproduce our new results, as well as the other multi-wavelength properties of these fascinating sources.
	
	\begin{acknowledgements}
		We thank the referee for the positive and constructive review of our work. We thank N.C. Stone and B.D. Metzger for very insightful discussions on the possible theoretical interpretations of our results. R.A. thanks K. Dennerl for discussions on the EPICpn response. We acknowledge the use of the matplotlib package \citep{Hunter2007:matplotlib}. G.P. acknowledges funding from the European Research Council (ERC) under the European Union’s Horizon 2020 research and innovation programme (Grant agreement No. [865637]). G.M. acknowledges funding from Project No. MDM-2017-0737 Unidad de Excelencia "Mar\'{i}a de Maeztu" - Centro de Astrobiolog\'{i}a (CSIC-INTA) by MCIN/AEI/10.13039/501100011033. MG is supported by the ``Programa de Atracci\'on de Talento'' of the Comunidad de Madrid, grant number 2018-T1/TIC-11733.
	\end{acknowledgements}

	%
	%
	\bibliographystyle{aa} 
	\bibliography{bibliography} 
	
	\begin{appendix}        
	\section{More details on the light curve fitting}
	\label{sec:xmm1}
	
	Obs1 requires multiple bursts to be fit. In Fig. A.1 we show the fit with three and four bursts. The latter improves on the former with a difference in 
	AIC values of $\Delta\text{AIC}_{4b,3b}\sim -167.8$. Residuals around $t - t_{Obs1,0}\sim20\,h$ (Fig. A.1 bottom) are significantly improved by adding a fifth burst (
	$\Delta\text{AIC}_{5b,4b}\sim -91.8$ with respect to the 4-bursts model). Therefore, we adopted the 5-bursts model as reference for Obs1 in the full energy band. We adopted the same procedure for Obs1 light curves extracted in small energy bins (see Fig.~\ref{fig:qpe1_edep_lc_xmm1}) to determine the number of bursts required. We report the comparisons between AIC values in Table~\ref{tab:logz_obs1_Edep}.
	
	\begin{table}[b]
		\small
		\caption{$\Delta$AIC values from the Obs1 3-, 4- and 5-bursts model runs.}
		\label{tab:logz_obs1_Edep}
		\centering
		\begin{tabular}{C{0.2\columnwidth} C{0.25\columnwidth} C{0.25\columnwidth}}%
			\multicolumn{1}{c}{Energy bin} &
			\multicolumn{1}{c}{$\Delta \text{AIC}_{4b,3b}$} &
			\multicolumn{1}{c}{$\Delta \text{AIC}_{5b,4b}$} \\           
			\midrule
			full & $-167.8$ & $-\textbf{91.8}$ \\
			$0.2-0.4\,$keV & $-41.6$ & $-\textbf{11.4}$ \\
			$0.4-0.6\,$keV & $-35.5$ & $-\textbf{26.9}$ \\
			$0.6-0.8\,$keV & $-\textbf{79.4}$ & $4.7$ \\
			$0.8-1.0\,$keV & $8.7$ & $-$ \\
			$1.0-2.0\,$keV & $7.0$ & $-$ \\
		\end{tabular}
		\tablefoot{$\Delta$AIC values are shown for the full energy range and for light curves extracted in different energy bins, as shown in the first column. Bold face highlights the best model adopted for each energy bin (with the lowest AIC value), which is the 3-bursts model otherwise.}
	\end{table}
	
\begin{figure}[tb]
		\centering
		\includegraphics[width=0.99\columnwidth]{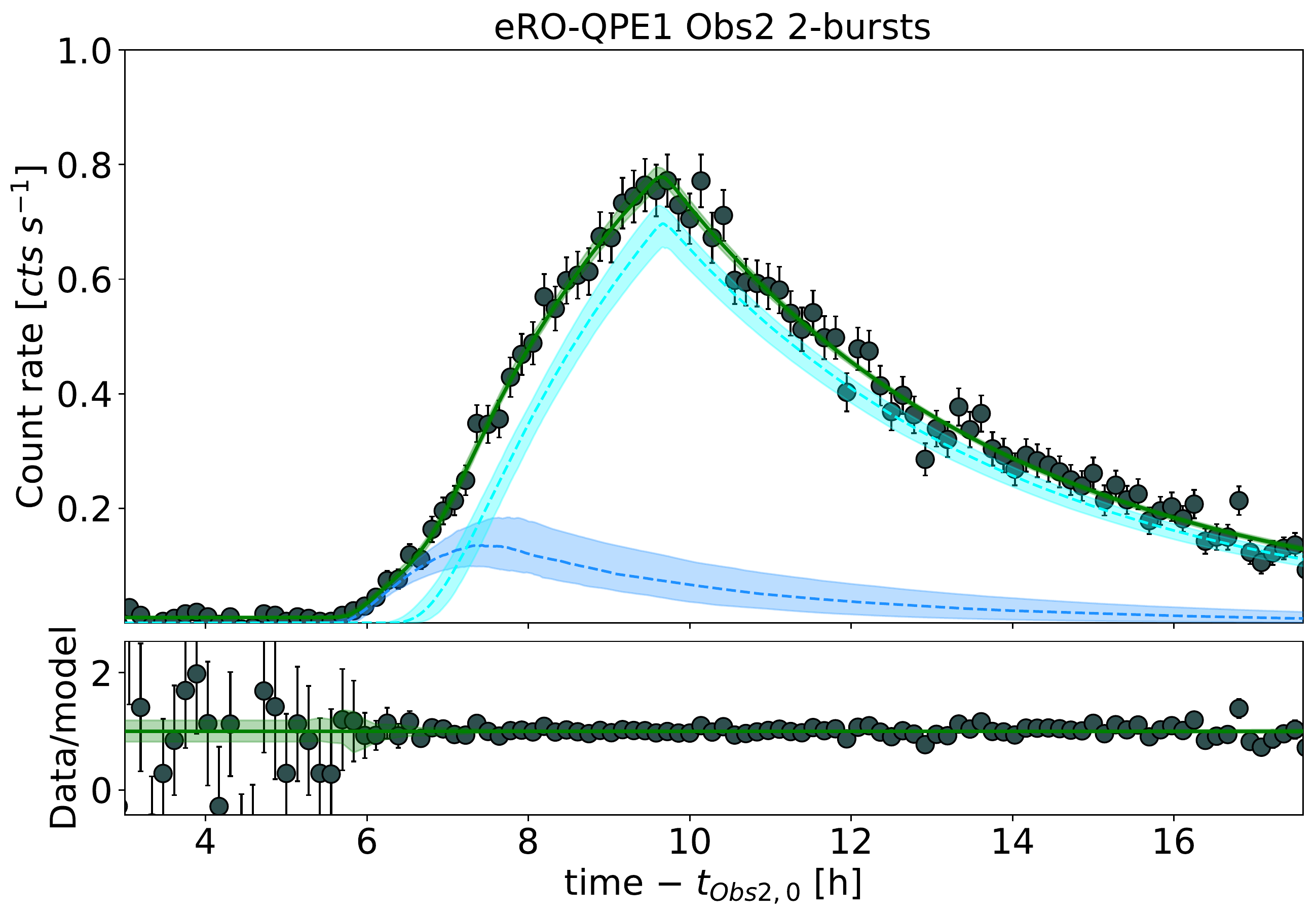}
		\caption{As in the right panel of Fig.~\ref{fig:qpe1_lc}, but fitting a model with two bursts (shown in cyan and light blue).
		}
		\label{fig:qpe1_lc_xmm2_2bursts}
	\end{figure}

	To further validate our results, we tested whether indeed a single burst profile is the preferred model for Obs2. We performed a fit with two overlapping bursts (see Fig.~\ref{fig:qpe1_lc_xmm2_2bursts}). Despite the good residuals, it is immediate to notice that one of the two bursts is relegated to fit a much smaller amount of counts and that Obs2 cannot be fit with two burst with compatible amplitude, which is instead the case for the first part of Obs1 (left panel of Fig.~\ref{fig:qpe1_lc}). Finally, comparing the AIC values of the single-burst and two-bursts fits ($\Delta\text{AIC}_{2b,1b}\sim5.5$), the former remains the preferred statistical description of Obs2 data. Hence, Obs2 is indeed best described with a single burst and since Obs1 requires many, the two observations of eRO-QPE1 do show a fundamental difference in timing properties, which occurred within just a few days and a handful of QPE cycles.
	
	\begin{figure}[tb]
		\centering
		\includegraphics[width=0.99\columnwidth]{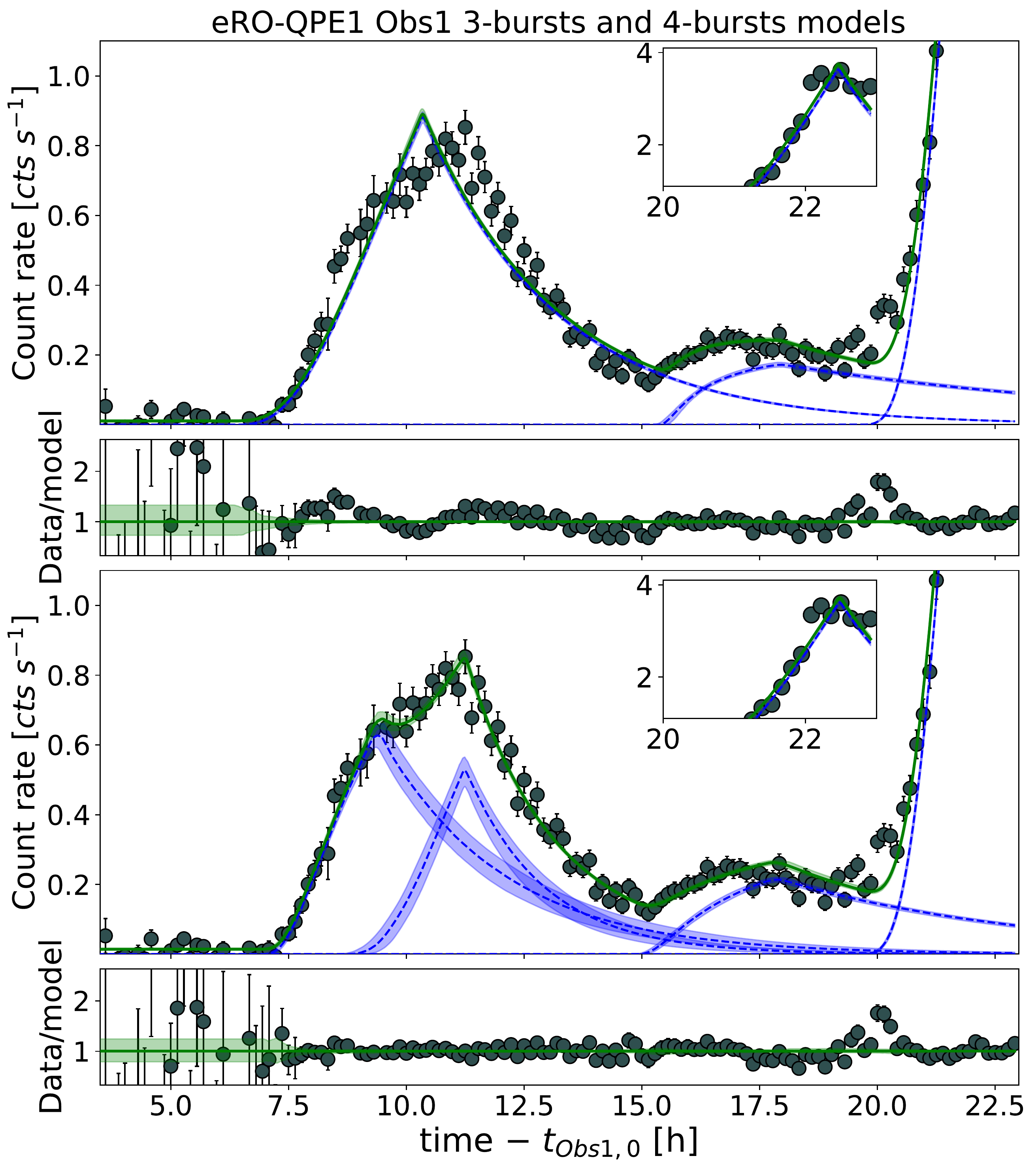}
		\caption{As in the left panel of Fig.~\ref{fig:qpe1_lc}, but fitting a model with three (top) and four (bottom) bursts.}
		\label{fig:qpe1_lc_xmm1_3and4}
	\end{figure}

\begin{figure}[tb]
		\centering
		\includegraphics[width=0.8\columnwidth]{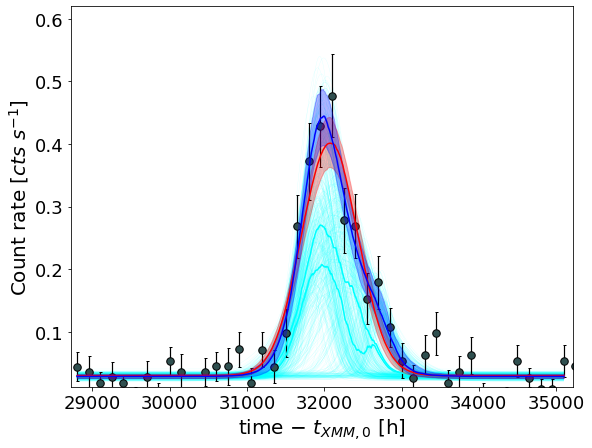}
		\caption{Eruption number four in the \emph{XMM-Newton} light curve of eRO-QPE2 (see also Fig.~\ref{fig:qpe2_lc}). The red curve represents the fit with one Gaussian component, in blue the superposition of two Gaussian profiles is shown, with the cyan lines showing the individual two Gaussian components (the median is shown with a thicker line).
		}
		\label{fig:qpe2_tests}
	\end{figure}

	We performed a similar test on the seemingly much simpler QPE source eRO-QPE2 (see Fig.~\ref{fig:qpe2_lc}, top). We fit the full light curve with one Gaussian and a superposition of two Gaussian profiles (obtaining AIC$_{2g}$ and AIC$_{1g}$, respectively). The simpler model is still preferred ($\Delta$AIC$_{2g, 1g}\sim10$). We also performed the same procedure for each single burst separately. In four bursts the AIC value of the two-Gaussian model is lower, while in the four remaining the single Gaussian is a good enough description of the data ($\Delta$AIC$_{2g, 1g}$ values of -4.6, -4.4, -3.4, 2.6, 1.5, -0.3, 3.9, 8.8, in order of time of the eruptions). However, as shown in Fig.~\ref{fig:qpe2_tests} with one example of burst with an improved fit, the analysis with two overlapping Gaussian profiles leads to multi-modal features covering the full parameters space of the burst (highlighted by the cyan lines), which is surely not physical. Motivated by the clear asymmetry in eRO-QPE1 we also tested the asymmetric model shown in Eq.~\ref{eq:1} on eRO-QPE2, obtaining for the full light curve an improvement of $\Delta$AIC$\sim-22$, with respect to the single Gaussian model. Indeed, the asymmetric model fit on the single bursts also yielded a compatible improvement as for the fit with the two-Gaussian model, when the latter was present in a single eruption. Clearly, dedicated analysis is needed for eRO-QPE2, which is beyond the scope of this paper and will be tested in future work. In contrast, we stress that in eRO-QPE1 both the asymmetry and the addition of multiple bursts are not only statistically significant, but also clearly evident in the data.

	\end{appendix}
	
\end{document}